\documentclass[twocolumn,noshowpacs,preprintnumbers,amsmath,amssymb,prb,superscriptaddress]{revtex4}
\usepackage{graphicx}
\usepackage{bm}
\usepackage{color}
\usepackage{textcomp}
\usepackage{gensymb}

\newcommand{\muchem}{\mu_{\mathrm{ch}}}

\begin{document}


\title{Magnetic-field driven ambipolar quantum Hall effect in epitaxial graphene close to the charge neutrality point}

\author{A. Nachawaty}
\affiliation{Laboratoire Charles Coulomb (L2C), UMR 5221 CNRS-Universit\'e de Montpellier, Montpellier, F-France.}
\affiliation{Laboratoire de Physique et Mod\'elisation (LPM), EDST, Lebanese University, Tripoli, Lebanon}

\author{M. Yang}
\affiliation{Laboratoire National des Champs Magn\'etiques Intenses (LNCMI-EMFL), UPR 3228, CNRS-UJF-UPS-INSA, 143 Avenue de Rangueil, 31400 Toulouse, France}

\author{W. Desrat}
\affiliation{Laboratoire Charles Coulomb (L2C), UMR 5221 CNRS-Universit\'e de Montpellier, Montpellier, F-France.}

\author{S. Nanot}
\affiliation{Laboratoire Charles Coulomb (L2C), UMR 5221 CNRS-Universit\'e de Montpellier, Montpellier, F-France.}

\author{B. Jabakhanji}
\affiliation{College of Engineering and Technology, American University of the Middle East, Egaila, Kuwait.}

\author{D. Kazazis}
\affiliation{Centre de Nanosciences et de Nanotechnologies, CNRS, Univ. Paris-Sud, Universit\'e Paris-Saclay, C2N – Marcoussis, 91460 Marcoussis, France}
\affiliation{Laboratory for Micro and Nanotechnology, Paul Scherrer Institute, 5232 Villigen-PSI, Switzerland.}

\author{R. Yakimova}
\affiliation{Department of Physics, Chemistry and Biology, Link\"oping University, SE-58183 Link\"oping, Sweden}

\author{A. Cresti}
\affiliation{Univ. Grenoble Alpes, CNRS, Grenoble INP, IMEP-LaHC, F-38000 Grenoble, France}

\author{W. Escoffier}
\affiliation{Laboratoire National des Champs Magn\'etiques Intenses, INSA UPS, CNRS UPR 3228,
	Universit\'e de Toulouse, 143 avenue de Rangueil, 31400 Toulouse, France}

\author{B. Jouault}
\affiliation{Laboratoire Charles Coulomb (L2C), UMR 5221 CNRS-Universit\'e de Montpellier, Montpellier, F-France.}

\begin{abstract}
We have investigated the disorder of epitaxial graphene close to the charge neutrality point (CNP)  by various methods: i) at room temperature, by analyzing  the dependence of the resistivity on the Hall coefficient  ; ii) by fitting the temperature  dependence of the Hall coefficient  down to liquid helium temperature; iii) by fitting the magnetoresistances at low temperature. All methods converge to give a disorder amplitude of $(20 \pm 10)$ meV.  
Because of this relatively low disorder, close to the CNP, at low temperature, the sample resistivity does not exhibit the standard value $\simeq h/4e^2$ but  diverges. Moreover, the magnetoresistance curves have a unique ambipolar behavior, which has been systematically observed for all studied samples. This is a signature of both asymmetry in the density of states and in-plane charge transfer. The microscopic origin of this behavior cannot be unambiguously determined. However, we propose a model in which the SiC substrate steps  qualitatively explain the ambipolar behavior.
\end{abstract}

\maketitle

\section{Introduction}

Undoubtedly, the best known exotic two-dimensional electron system is graphene.\cite{AKGeim2007}
Among various exciting properties, this material has demonstrated a half-integer quantum Hall effect~\cite{Zhang2005} (QHE) which is very robust in temperature,\cite{Novoselov2007} because of the extremely large energy separation between the first Landau levels (LL) lying close to the bottom of conduction and valence bands.
%

The QHE in graphene is strongly influenced by disorder and hence 
by the choice of the substrate.
The quantum Hall plateaus observed in graphene on SiC (G/SiC) have 
a high breakdown current\cite{Alexander-Webber2013}
and appear at lower magnetic fields\cite{Huang2015}
with respect to graphene deposited on SiO$_2$.\cite{Poumirol2010}
The quantum plateaus in G/SiC devices are much larger in magnetic field than
those obtained in graphene encapsulated in hBN,\cite{Dean2011} because they are
stabilized by charge transfer\cite{Janssen2011} and disorder.\cite{Lafont2015}%
Thanks to these properties, it was recently demonstrated that G/SiC can act as a quantum electrical resistance standard,\cite{Tzalenchuk2010} even in  experimental conditions relaxed with respect to the state-of-the-art in GaAs-based quantum wells.\cite{Ribeiro-Palau2015}

To date, fundamental and practical questions remain open.
In particular, the fate of the G/SiC quantum plateaus close to the charge neutrality point (CNP) has still to be elucidated. 
Achieving and controlling low doping is of primary interest
 to use graphene as a resistance standard at even lower magnetic fields, or in cryogen-free systems.\cite{Janssen2015} G/SiC could be also a material of choice for testing theoretical models predicting additional quantum plateaus depending on the type of disorder.\cite{Nomura2008,Ostrovsky2008}

\begin{figure}
	\includegraphics[width=0.95 \linewidth]{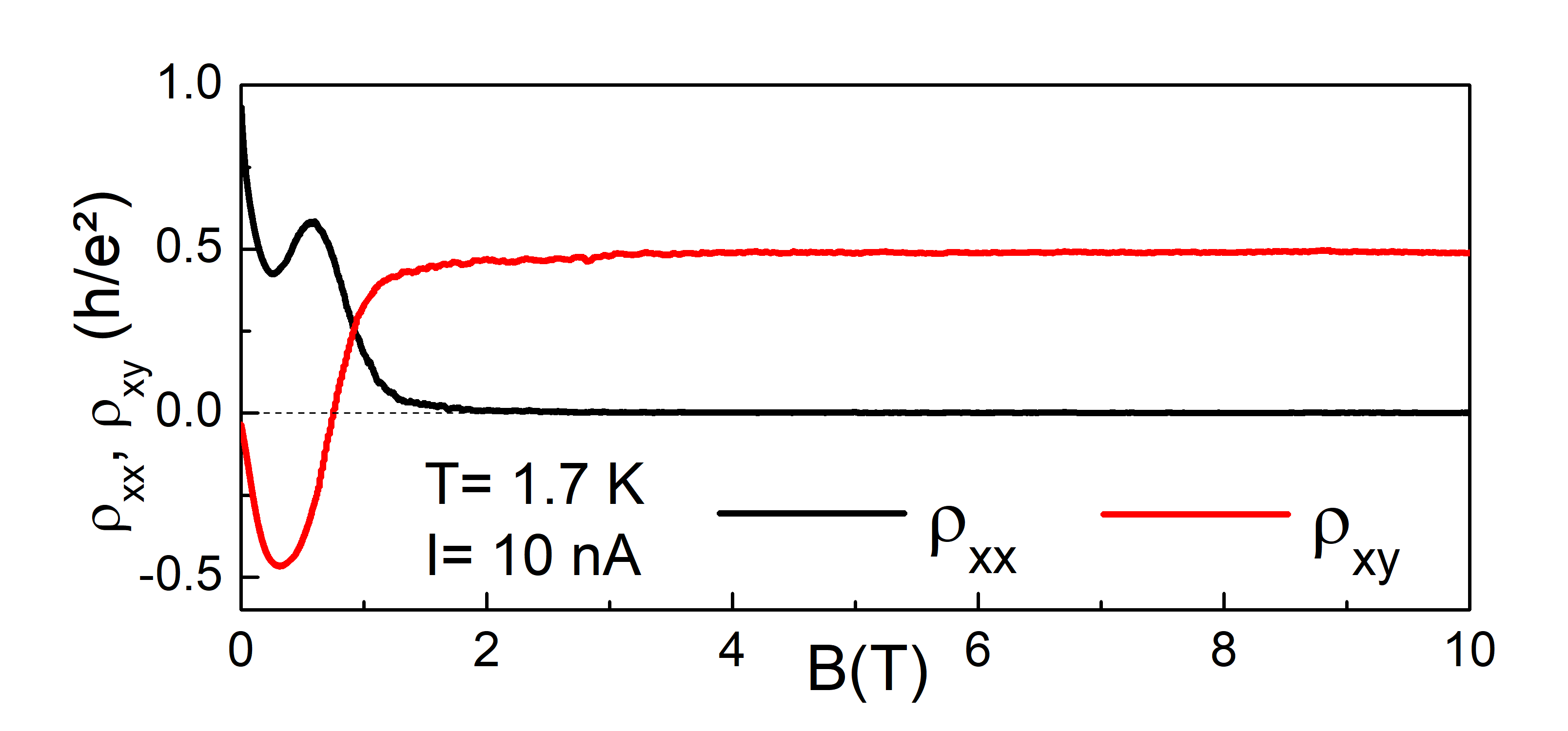}
	\caption{Magnetoresistance observed in sample G31 at low temperature $T=1.7$ K and low current $I= 10$ nA. A quantum Hall plateau at $\rho_{xy}=R_K/2$ initially develops when $B$ increases up to  0.5~T. At higher $B$, the Hall resistance changes sign and a long plateau appears at $\rho_{xy}= R_K/2$. The longitudinal resistance $\rho_{xx}$ shows a pronounced maximum when the Hall resistance changes sign. $\rho_{xy}<0$ ($>0$) corresponds	to holes (electrons).}
	\label{fig:AQHE}
\end{figure}

However, there exist only a few  experimental analyses of the transport properties close to the CNP,\cite{Huang2015}
mainly because G/SiC is intrinsically strongly $n$-doped by the SiC substrate,
thus requiring the system to be compensated by a top gate.
For graphene on SiO$_2$, close to the CNP, QHE reveals that electrons and holes coexist even when the energy spectrum is quantized and the carriers partially localized,\cite{Wiedmann2011,Kurganova2013} but for G/SiC 
even this rather intuitive picture has not been thoroughly tested. The amplitude of the disorder potential fluctuation has been evaluated by various methods,\cite{Curtin2011,Huang2015}
 but the type of disorder and its spatial and energy distribution close to CNP remain mostly unknown.

In this paper, we present magnetotransport experiments in G/SiC to evaluate  the disorder and to test the stability of the QHE close the CNP. 
The results are largely unexpected and reveal new physics in comparison to 
what
is observed on other substrates.
We show that not only electrons and holes coexist in high magnetic fields, but the carriers also redistribute unexpectedly as a function of magnetic field $B$, as illustrated in Fig.~\ref{fig:AQHE}. 
In this figure, the graphene is tuned very close to the CNP. The  magnetoresistance reveals a magnetic field driven ambipolar QHE. At low $B$, 
a quantized $p$-like plateau starts to develop. However, when $B$ is increased further, 
this plateau collapses and is replaced by a quantized $n$-like plateau of opposite sign. We show that this behavior is robust and reproducible for the range of doping where both types of carriers coexist and we propose a model based on disorder and in-plane charge transfer, after a detailed analysis of the sample disorder close to the CNP.

\section{Methods and methodology}

\subsection{Sample fabrication}
The SiC/G samples have been grown epitaxially on the Si-face of a semi-insulating 4H-SiC substrate at a high temperature $T= 2000$\degree{C}. The as-grown samples have large uniform monolayer areas.
Atomic force microscope analysis revealed the presence of SiC steps, 
approximately 500 nm wide and 2 nm high, uniformly covered by the graphene layer. Additional Raman analysis revealed the presence of elongated bilayer graphene patches, approximately   10 {\micro m}
 long and  2 {\micro m}
 wide, covering around 5\% of the total surface.

Hall bars of various size and geometry were then fabricated by standard electron-beam lithography. The graphene was covered by two layers of resist, as described in Ref.~\onlinecite{Lara-Avila2011}.
The resist acts as a chemical gate and strongly reduces the intrinsic carrier density of the graphene layer.

Magnetotransport measurements were performed on four SiC/G Hall bars, named G14, G31, G21 and G34. 
The bars have a length of 420 {\micro m}
 and a width of  100 {\micro m},
except for G34, which has a width of 20 {\micro m}. 
%


\subsection{Corona preparation}

\begin{figure}
	\includegraphics[width=0.95 \linewidth]{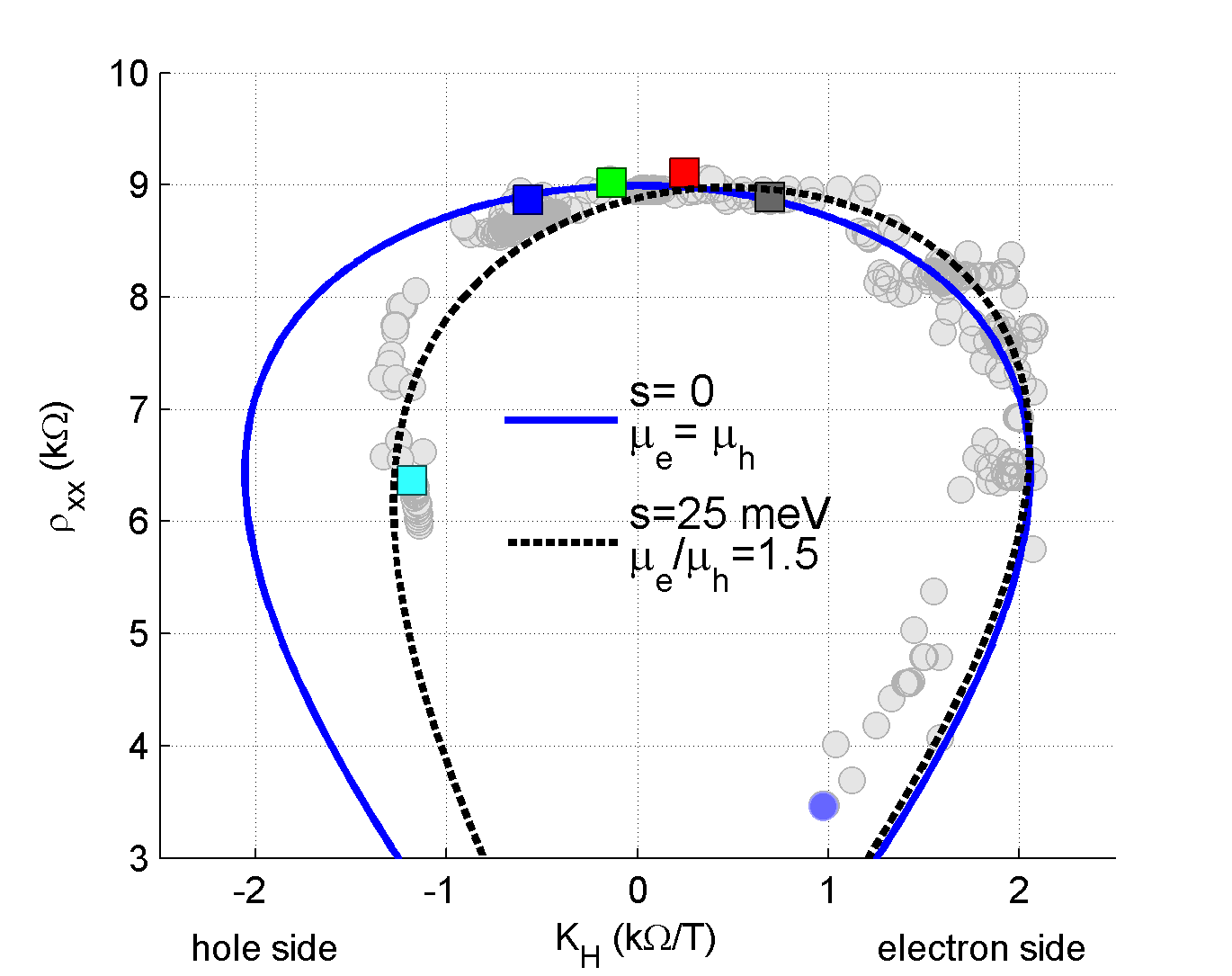}
	\caption{$\rho_{xx}$ {\it vs} $K_H$ during the corona preparation for sample G14. The initial state before the ion deposition is indicated by the blue filled circle.  The colored squares are various states from which the G14 sample has been cooled down at low temperature, see Fig.~\ref{fig:MR}a. The blue line is a fit without disorder ($\mu_e=\mu_h= 4,300$ cm$^2$/Vs), the black line a fit where the disorder is taken into account  ($s=25$ meV, $\mu_h= 2,750$ cm$^2$/Vs, $\mu_e/\mu_h= 1.5)$.}
	\label{fig:circle}
\end{figure}

After the lithography process,  the graphene layer was systematically
$n$-doped with $n \simeq 6 \times 10^{11}$ cm$^{-2}$ at room temperature.
Carrier density control was performed with ion projections onto the resist bilayer covering graphene. Negative ions were produced by repeated corona discharges with a time interval of 17~s, following the method described in Ref.~\onlinecite{Lartsev2014}. The distance between the sample and the corona source was 12~mm. Changes in the electronic properties of graphene upon exposure to corona ions were detected by continuous measurements of the resistance 
and  Hall coefficient $K_H$ at room temperature, using low magnetic field ($B$= 0.05 T) and dc current $I= \pm 1$ {\micro A}. The evolution of the resistivity $\rho_{xx}$ as a function of the Hall coefficient $K_H$ during exposure to ions is presented in Fig.~\ref{fig:circle}. 
There is some point dispersion due to the absence of correlation between the discharges and the electrical measurements.
However, the carrier density clearly changes from its initial $n$-doping of $6 \times 10^{11}$ cm$^{-2}$ to a $p$-doping of 
$5 \times 10^{11}$ cm$^{-2}$ after a few hundreds cycles of corona discharge.
After this point, additional ion projections are inefficient to increase further the $p$-doping. When the discharges are stopped,
the carrier density drifts slowly towards $n$-doping and $K_H$
stabilizes around  $+ 1 $ k$\ohm$/T within a few hours. 
The initial carrier density is not recovered even when the sample is left several months under ambient atmosphere.

\section{Estimation of disorder}

\subsection{Disorder estimated from $K_H(\rho_{xx})$ at room temperature}

To describe the sample evolution seen in Fig.~\ref{fig:circle}, we use the usual equations~\cite{Li2011} that give the conduction of a homogeneous sample, in which both electrons and holes participate in the conduction in parallel because of thermal activation.
The model is detailed in Annex.
The modeled $(\rho_{xx}, K_H$) curve is plotted for $\mu_e=\mu_h= 4,300$ cm$^2$/Vs and $T= 300$ K as a blue solid line in Fig.~\ref{fig:circle}a when 
$\muchem$ spans the energy window around the CNP.
Here, $\mu_e$, $\mu_h$ and $\muchem$
are the electron mobility, hole mobility and chemical potential respectively.
The model fits fairly well the data but with an obvious deviation on the hole side, where the $K_H$ coefficient is overestimated. 
The asymmetry
of the $(\rho_{xx}, K_H$) curve indicates that
$\mu_e$ and $\mu_h$ differ.
The data can be modeled more precisely in two ways:
i) the mobility ratio $\mu_e/\mu_h$ increases when the number of deposited ions increases because 
negative ions have a larger cross section for $p$-type charge carriers and their presence can decrease significantly the hole mobility~\cite{Novikov2007}; 
ii) $\mu_e/ \mu_h > 1$ and does not depend on the charge carrier concentration but disorder has to be taken into account.

\begin{figure}
	\includegraphics[width=0.999 \linewidth]{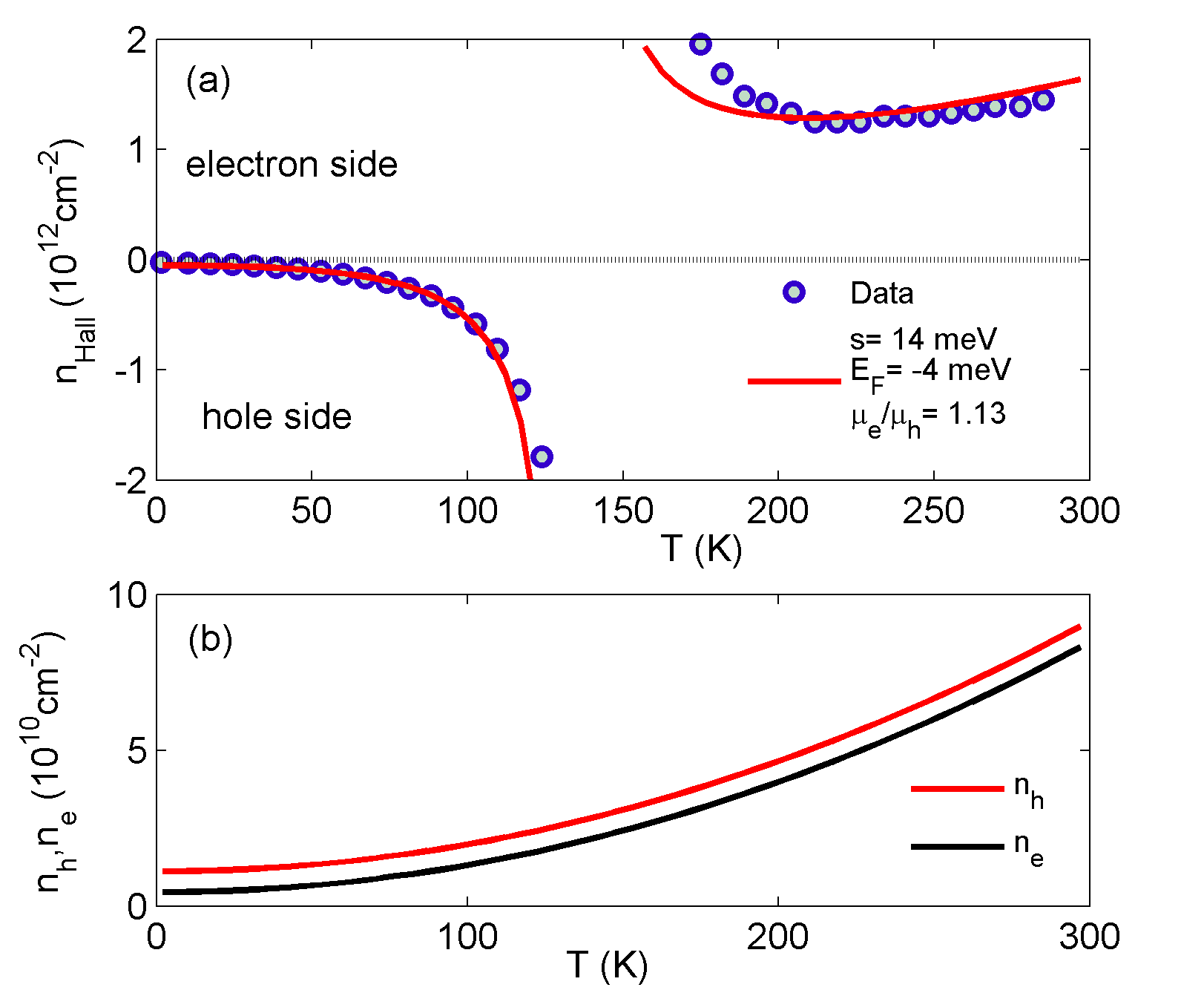}
	\caption{(a) Temperature dependence of the Hall carrier density $n_\mathrm{Hall}= 1/(\rho_{xy}e)$ for sample G14 (open circles). A good fit (red solid curve) is obtained by including disorder ($s=14$ meV) with two other fitting parameters:  the Fermi energy $E_F$=-4 meV at $T=0$ K and the mobility ratio $\mu_e/\mu_h$= 1.13. By taking $s=0$, the experimental data cannot be fitted properly. (b) electron and hole density {\it vs} $T$  given by the best fit in (a) are indicated  by black and red solid lines.}
	\label{fig:RT}
\end{figure}

With the disorder amplitude $s$ as an additional fitting parameter,
it is possible to fit the asymmetry of the $\rho_{xx}(K_H)$ curve, as
shown by the black line in Fig.~\ref{fig:circle}. The fitting parameters are $s= $25 meV, $\mu_e/\mu_h$= 1.5 and $\mu _h$= 2,750 cm$^2$/Vs. All these parameters are in agreement with the literature.~\cite{Huang2015}

\subsection{Disorder estimated from $n_\mathrm{Hall}(T)$}
%
The disorder can also be estimated from the temperature dependence of the Hall coefficient. 
Fig.~\ref{fig:RT}a shows the evolution of the Hall density
$n_\mathrm{Hall}= 1/(K_H e)$ of sample G14 when $T$ is lowered from room temperature down to 1.7 K. The initial corona preparation has been chosen to illustrate
that one should not expect a $T^2$ dependence of the Hall density as a universal trend,
as assumed in Ref.~\onlinecite{Huang2015}.
In Fig.~\ref{fig:RT}a, 
the Hall density diverges at a critical temperature $T_c \simeq 140$ K,
is negative below $T_c$ and positive above $T_c$.
This divergence is a clear signature that both electrons and holes participate in the conduction and
from Eq.~\ref{eq:rhoxy}, $T_c$ can be identified as the temperature for which
$n_h \mu_h^2= n_e\mu_e^2$.
We stress that this change of sign of $n_\mathrm{Hall}$ does not imply that the net carrier density changes with $T$.
Let us assume, as in Refs.~\onlinecite{Li2011} and \onlinecite{Huang2015}, that the net carrier density
$n= n_e-n_h$ does not depend on $T$.
It is possible to fit the data taking into account 3 parameters: the disorder amplitude $s$, 
the Fermi energy at zero temperature $E_F= \lim_{T\rightarrow 0} \muchem$,
 and the mobility ratio $\mu_e/\mu_h$.
From the fit of the temperature dependence shown in Fig.~\ref{fig:RT}a, we extract $s= 14$ meV, $E_F$ = -4 meV and $\mu_e/\mu_h= 1.13$.
The fit matches very well the data, except in the vicinity of $T_c$, where the experimental error is larger as the measured voltage cancels.
Fig.~\ref{fig:RT}b shows that $n_e$ and $n_h$ increase quadratically with $T$.

From various $n_\mathrm{Hall}(T)$ fits
corresponding to 11 different initial corona preparations, 
we found 
$s= (21 \pm 12)$ meV and  $\mu_e/\mu_h = 1.0 \pm 0.15$.
These parameters are roughly in agreement with those extracted from the $K_H(\rho_{xx})$ fit.
However, the precision is not good enough to evidence 
a possible evolution of $\mu_e/\mu_h$ with the progressive ion deposition.

\section{Ambipolar quantum Hall effect}

\begin{figure*}
	\includegraphics[width=0.95 \linewidth]{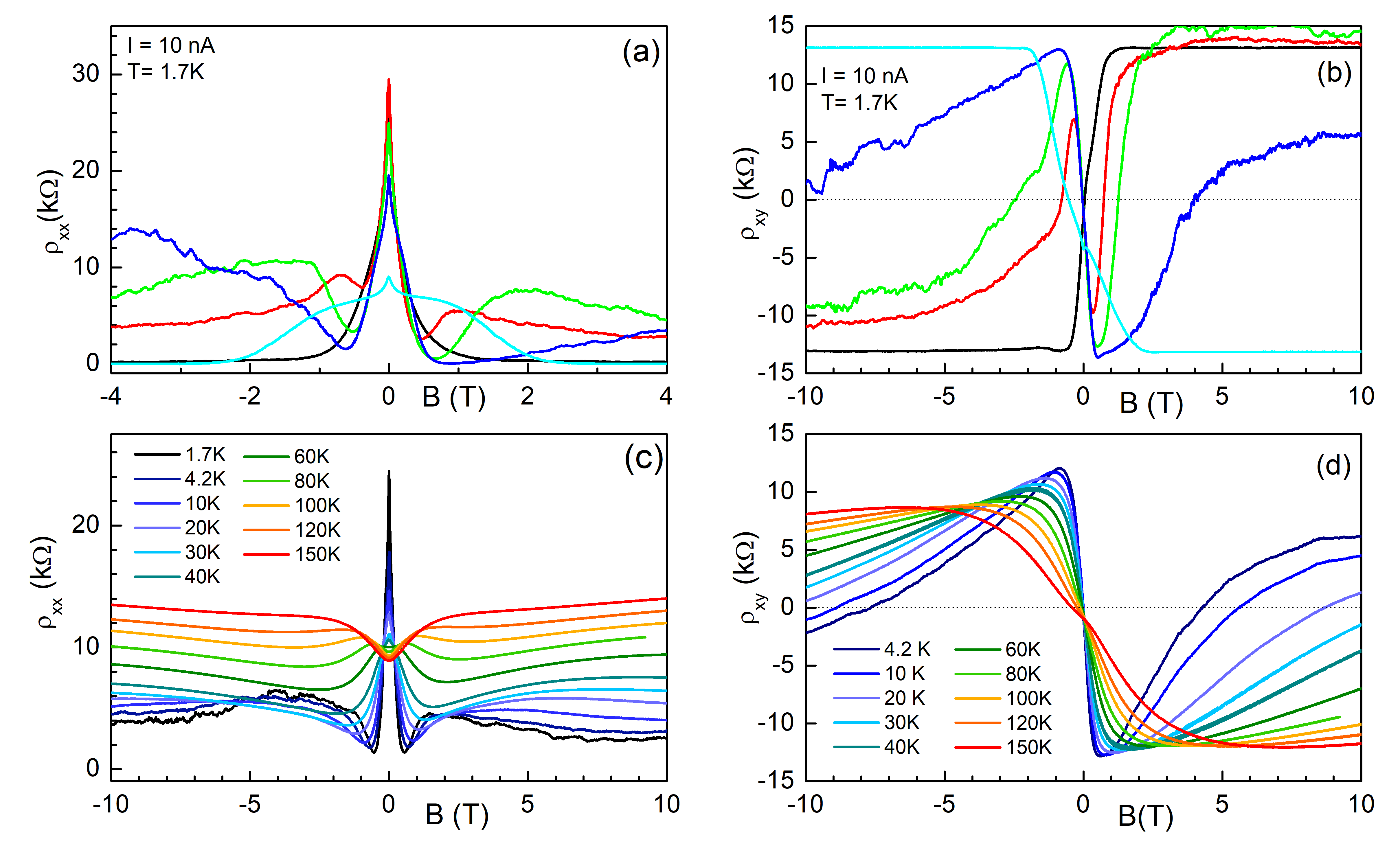}
	\caption{(a) Longitudinal and (b) transverse magnetoresistances observed in sample G14 at $T=1.7$ K and $I= 10$ nA. The different curves correspond to five different sample preparations, indicated by the colored squares in Fig.~\ref{fig:circle}. (c,d) Temperature dependence of the longitudinal and Hall magnetoresistances ($\rho_{xx}$ and $\rho_{xy}$) for an additional sample preparation, close to the red square in Fig.~\ref{fig:circle}.}
	\label{fig:MR}
\end{figure*}

\subsection{Experimental results}

Fig.~\ref{fig:MR}a,b shows the longitudinal and transverse magnetoresistances
at $T= 1.7$ K for sample G14.  
Five different dopings, indicated in Fig.~\ref{fig:circle} by colored squares, 
were first obtained at room temperature by the corona method. For
each of these initial preparations, 
the sample was introduced in the cryostat and cooled down.


We first note that the longitudinal resistance is unusual.
There is a pronounced negative magnetoresistance peak around $B=0$ T, see Fig.~\ref{fig:MR}a.
The peak value is close to $h/e^2$ at $T$ = 1.7 K, and largely exceeds the pseudo-universal\cite{Adam2007} value of $\rho_{xx} \simeq h/4e^2$ usually observed in graphene on SiO$_2$ close to the CNP. We checked that $\rho_{xx}$ increases even further, up to $\simeq 70$ k$\Omega$, when the temperature is decreased down to 280 mK. This insulating behavior, see Fig.~\ref{fig:MR}c,  shares strong similarities with a previous experiment,\cite{Ponomarenko2011} 
where an insulating behavior observed in graphene encapsulated in hBN
was attributed to Anderson localization. Following Ref.~\onlinecite{Ponomarenko2011},
such a localization can only be observed if the carrier density in the puddles is small enough (around $10^{10}$ cm$^{-2}$ in Ref.~\onlinecite{Ponomarenko2011}), 
in good agreement with our own estimation of the disorder in G/SiC
($s \simeq 25$ meV yields $n \simeq 1.4\times 10^{10}$ cm$^{-2}$.)  


We now focus on the Hall magnetoresistance. First, the usual half-integer QHE
is observed for high $p$-doping ($n_h \simeq 10^{11}$ cm$^{-2}$, panel b, cyan curve)
or low $n$-doping ($n_e \simeq 10^{10}$ cm$^{-2}$, panel b, black curve).
%
Between these two dopings, the plateaus are not well defined. However,  the
Hall magnetoresistance  systematically follows a remarkable behavior.
Increasing $B$, there is first a 
decrease of $\rho_{xy}$,
followed by a saturation at or before $\rho_{xy}= -R_K/2$
($p$-like plateau), where $R_K=h/e^2$. 
Then, $\rho_{xy}$ collapses, 
changes its sign 
and finally stabilizes at a positive value close to $R_K/2$ ($n$-like plateau).
Using pulsed magnetic fields, we checked that there is no more change of sign of $\rho_{xy}$ at least up to 30 T.
We never observed the opposite transition, from $n$-like to $p$-like plateau.

In some other measurements, as shown in Fig.~\ref{fig:AQHE}, a well defined plateau has been observed at $R_K/2$. 
The magnetic field at which $\rho_{xy}$ cancels is also related with the  appearance of a bump on 
$\rho_{xx}$, see Fig.~\ref{fig:AQHE} and Fig.~\ref{fig:MR}a. 
The position of this bump seems to be also related with the initial doping obtained at room temperature. When graphene is slightly $p$-doped at room temperature, this $\rho_{xx}$ bump appears at low $B$ (red curve in panel a). When graphene becomes progressively more $p$-doped, the bump shifts to higher $B$ (green and blue curves in panel a) and finally disappears (cyan curve).

Finally both $\rho_{xx}$ and $\rho_{xy}$ show a clear temperature dependence, as shown in panels c,d. The additional $\rho_{xx}$ bump goes to higher $B$ when $T$ increases and then disappears. The $\rho_{xy}$ behavior is even more striking, as the ambipolar behavior disappears above $T \simeq 20$ K and is replaced by an almost quantized plateau corresponding  to an apparent $p$-doping.


\subsection{Inadequacy of the standard two fluid model}
The two fluid Drude model   (Eqs.~\ref{eq:rhoxy} and~\ref{eq:rhoxx}) 
can explain in some cases a change of sign of $\rho_{xy}$.
This is because at low $B$, 
$\rho_{xy} \simeq (n_e\mu_e^2-n_h\mu_h^2)/ (n_h\mu_h+n_e\mu_e)^2$
whereas at high $B$,
$\rho_{xy} \simeq 1 / (n_e- n_h)$.
However,
Fig.~\ref{fig:MR}d shows that in the high field limit $B \simeq 10$~T,
$\rho_{xy}$ is positive at low $T$.
This is in contradiction with the previous temperature analysis, which indicates
that  $n_e-n_h < 0$ at all $T$.
Moreover $\rho_{xy}$ is negative at low $B$, which,
in the framework of two fluid Drude model, indicates that holes
are more mobile than electrons, in contradiction with all our previous analyses.
Additionally,  the Drude model is not valid in the quantum Hall regime studied here
and
predicts only monotonous positive longitudinal magnetoresistance,
in contradiction with the observation of a  bump in $\rho_{xx}$ at finite $B$.
Obviously, the two fluid model fails to explain the magnetoresistances and another explanation is needed to account for all of these observations.

%

\begin{figure}
	\includegraphics[width=0.95 \linewidth]{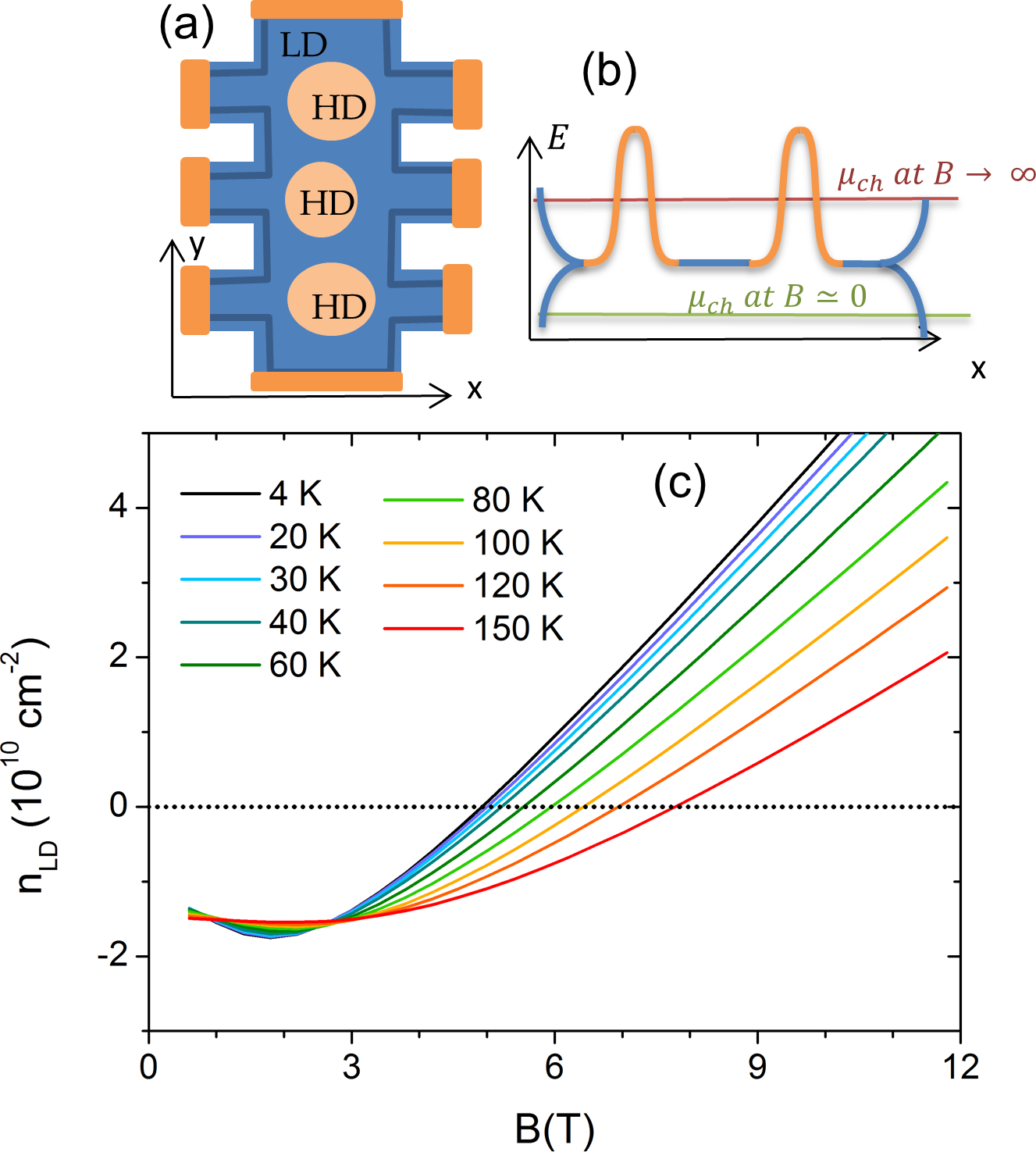}
	\caption{(a) Top-view sketch of the Hall bar, embedding regions (HD, orange color) whose CNP is at higher energy than the surrounding graphene region (LD, blue color). The edge states of the devices are also presented as dark blue lines. (b) Profile of the LL near the CNP. The LD and HD regions are indicated by blue and orange lines respectively. The LL splits at the sample edges.  When $B$ increases, the electrochemical potential tries to maintain a constant total charge and may cross the CNP of the LD region at a finite $B$. (c) Estimated carrier density $n_{LD}$ in the LD region {\it vs} $B$, for the model given in the main text,$\mu_\mathrm{CNP}^\mathrm{LD}= 14$ meV, $\Gamma=15$ meV, $\alpha$ = 30 \%.}
	\label{fig:model}
\end{figure}

\subsection{Disorder-induced charge transfer model}

A rather similar ambipolar behavior was reported in graphene on  SiO$_2$ substrate and interpreted as being due to important disorder.\cite{Poumirol2010} Here,
$\rho_{xy}$ changes sign at much smaller $B$ and the QHE is rather well preserved, suggesting that the disorder amplitude is much smaller.

Besides, charge transfer has already been identified as being a major source of charge redistribution
in G/SiC when the magnetic field evolves.\cite{Janssen2011}
The SiC interface states and graphene are coupled by a quantum capacitance, which depends on the graphene density of states (DOS) and hence on $B$.\cite{Kopylov2010} Additional in-plane charge transfers, induced by disorder, have also been identified in G/SiC when the overall top gate is degraded.\cite{Yang2016}
Here, the photochemical top gates have kept their integrity (as routinely observed optically). Nevertheless,
 we show in the following that in-plane charge transfer  explain 
the observed ambipolar QHE.
%

Let us assume that the main part of the Hall bar is homogeneous, with a low and uniform $p$-doping. This region is referred to as the low doped (LD) region. 
However, the device also contains a few highly doped (HD) regions where the $p$-doping is higher (Fig.~\ref{fig:model}a).
%
The key ingredient of the model is that, in the quantum regime, the conductivity is governed by the edge states of the LD region. The HD regions do not percolate, do not participate in the conduction and only act as additional charge reservoirs. 

The possible microscopic origins of these regions are multiple.
For instance, the role  of ionized acceptors in a quantizing magnetic field is well documented in the literature.\cite{Kubisa1996}
They trap electrons, create an impurity band in the DOS of the
LLs at high energies and shift the position of the quantum plateaus in two-dimensional electron gases.\cite{Haug1987}
In our case,
the ions trapped in the resist could act as magneto-acceptors, but we
have no definitive proof for this scenario.

In what follows, we propose another microscopic origin for the HD regions. 
It is well known that the SiC surface is not flat but has
a step-like structure.
The graphene layer is lifted from the SiC surface close to the SiC step edges.
There, the quantum capacitance between graphene and the SiC interface is reduced, leading to a larger $p$-doping.\cite{Low2012}

At low magnetic field, graphene can be prepared in such a way that 
$\muchem$ is below the delocalized states of the LD region
and the conductivity is governed by edge states which have a hole character, as illustrated in Fig.~\ref{fig:model}b (green line).
When the magnetic field is increased, the LL degeneracy increases.
As the total net carrier density (including LD and HD regions) tends to remain constant, $\muchem$ moves to higher energy (red line).
When the magnetic field is high enough, $\muchem$ has shifted between the LL energies of the LD and HD regions. There, the LD region has become $n$-doped whereas the HD puddles are still $p$-doped. The conductivity is then governed by the LD edge states which have now an electron character. 
Experimentally, the magnetic field at which $\rho_{xy}$ reverses can be as low as $B_m= 0.5$ T, which gives the minimal size of the HD regions:
$\sqrt{\hbar/eB_m} \simeq $ 40 nm.

Below we derive numerical estimates for this model assuming that the inhomogeneity
comes from the step-like structure.
First, because of the quantum capacitance taking place between 
 states  at the G/SiC interface and graphene~\cite{Kopylov2010}, 
any modification of the DOS structure due to the magnetic field induces charge displacement.
For each  region LD and HD, the balance equation gives:
\begin{equation}
n^i(\muchem)= -n_g + \beta^i (A + \mu_\mathrm{CNP}^{i} - \muchem)
\end{equation}
where $i$ labels the LD and HD regions, 
$n_g$ is the gate charge density,
$A$ is the workfunction difference between undoped graphene and the interface states, 
$\beta^i$ is an effective density of states, 
$\mu_\mathrm{CNP}^{i}$ is the potential of the CNP in region $i$.
We choose $\mu_\mathrm{CNP}^\mathrm{LD}=0$ as a reference.
$\beta^i$ is given by 
$\beta^i=  \epsilon_0 / (\epsilon_0+ e^2d^i\gamma)$
where $\epsilon_0$ is the vacuum dielectric constant, 
$\gamma$ is the density of interface states,
$d^i$ is the distance between graphene and the interface.
 
 At equilibrium, the electro-chemical potential is the same in the whole sample.
Following Ref.~\onlinecite{Yang2016},
the equation to solve is then obtained by summing the contributions of $n^\mathrm{HD}$
 and $n^\mathrm{LD}$:
 \begin{align}
 	(1-\alpha) n^\mathrm{LD}(\muchem) + \alpha n^\mathrm{HD}(\muchem) = - n_g  \nonumber \\ 
 	+(1-\alpha) \beta^\mathrm{LD}\left( A - \muchem\right)  \nonumber \\
 	+\alpha \beta^\mathrm{HD} \left( A + \mu_\mathrm{CNP}^\mathrm{HD}- \muchem\right),
 	\label{eq:transfert}
 \end{align}
where $\alpha$ is the proportion of HD region.
 On the left side of this equation, 
 both $n^\mathrm{LD}$ and  $n^\mathrm{HD}$ can be numerically estimated at a given magnetic field, where the DOS is given by a sum of Landau levels, each with a fixed Gaussian broadening $\Gamma$.
 %
 The chosen parameters are
 $n_g$ = $1.59 \times 10^{10}$ cm$^{-2}$, $A$= 0.4 eV, 
 $\gamma$ =$5 \times 10^{12}$ cm$^{-2}$eV$^{-1}$,
 $\Gamma$= 15 meV.
 They have been chosen in accordance with the literature.\cite{Janssen2011}
We assume that the LD and HD regions only differ
by a very small difference between $d^\mathrm{LD}$ and $d^\mathrm{HD}$:
$d^\mathrm{LD}$= 0.3 nm and $d^\mathrm{HD}= 0.4$ nm.
This small difference induces a shift
between the CNPs of the two LD and HD regions which is comparable to 
our estimation of the disorder:
$\mu_\mathrm{CNP}^\mathrm{HD}$ = 14 meV $\simeq s$.

Fig.~\ref{fig:model}c shows $n^\mathrm{LD}(B)$ obtained from solving Eq.~\ref{eq:transfert}, with 
$\alpha =30$\%.
As the model assumes that the conduction is governed by the LD region alone, 
the magnetic field dependence of $n^\mathrm{LD}(B)$
reproduces the ambipolar behavior of the transverse magnetoresistance.
Note that a qualitative agreement of the $T$-dependence is also obtained.
At low $B$, $n^\mathrm{LD}$ remains essentially $T$-independent, in agreement
with the previous $n_\mathrm{Hall}(T)$ analyses. By contrast,
the critical magnetic field at which the sign of $\rho_{xy}$ 
({\it i.e.} the sign of $n_\mathrm{LD}$) reverses is shifted to higher $B$ when $T$ increases. Experimentally this trend is indeed observed, with a shift which is even more pronounced, suggesting that some other parameters ({\it e.g.} $A$ or $\gamma$) could also be $T$-dependent.

Finally, this model relates unambiguously  the $\rho_{xx}$ bump 
to the conduction through the delocalized states of the LD region.
The model also predicts that this bump shifts to higher $B$ when the initial $p$-doping increases, as indeed experimentally observed.

\section{Conclusion}
To conclude, we have investigated the disorder of epitaxial graphene close to the charge neutrality point  by various analyses of the transport properties. All these analyses converge to give a disorder amplitude of the order of a few tens of meV.  Remarkably , the magnetoresistance curves have an ambipolar behavior  driven by the magnetic field. We interpret this as the signature of a very specific disorder combined with in-plane charge transfer between different regions in the graphene layer. The origin of disorder cannot be unambiguously determined but numerical estimations show that it could be related to the stepped SiC substrate.

\section*{Acknowledgment}
We thank 
F. Schopfer and W. Poirier (Laboratoire national de m\'etrologie et d'essais, France) for fruitful discussion. This work has been supported in part by the French Agence Nationale pour la Recherche  (ANR-16-CE09-0016) and by Programme Investissements d'Avenir under the program ANR-11-IDEX-0002-02, reference ANR-10-LABX-0037-NEXT. Part of this work was performed at LNCMI under EMFL proposal TSC06-116.

\section*{Annex: Model of disorder}
To describe the sample evolution seen in Fig.~\ref{fig:circle}, we reproduce below the usual equations~\cite{Li2011} that give the conduction of a homogeneous sample, in which both electrons and holes participate in the conduction in parallel because of thermal activation.
The total electron density is given by:
\begin{equation}
n_{e} =  \int_{-\infty}^{\infty} D_{e} (E) f(E - \muchem) dE,
\label{eq:ne}
\end{equation}
$D_{e}(E)=  D_1 E \theta(E)$ is the density of states (DOS) for electrons,
$D_1= g_s g_v/ (2\pi \hbar^2 v_F^2)$,
$g_s=2$ and $g_v=2$ are the spin and valley degeneracies,
$f= 1/\left[1+\exp((E-\muchem)/k_BT)\right]$ is the Fermi distribution function,
$\theta$ is the Heavyside function
and $\muchem$ is the chemical potential.
The total hole density is given by:
\begin{equation}
n_{h} =  \int_{-\infty}^{\infty} D_{h} (E) (1-f(E - \muchem)) dE,
\label{eq:nh}
\end{equation}
where $D_h(E)= D_e(-E)$ is the DOS for holes.
At low magnetic fields $\mu_e B, \mu_h B \ll 1$,
the Hall and longitudinal resistivities are given by:
\begin{align}
\rho_{xy}= -\rho_{yx}=  \nonumber \\ - \frac{1}{e} 
\frac{(n_h\mu_h^2-n_e\mu_e^2)+ \mu_h^2\mu_e^2B^2 (n_h-n_e)}
{(n_h\mu_h+n_e\mu_e)^2 + \mu_h^2\mu_e^2 (n_h-n_e)^2B^2}
B
\label{eq:rhoxy}
\end{align}
and 
\begin{equation}
\rho_{xx}= \frac{1}{e} 
\frac{n_h\mu_h+n_e\mu_e+ (n_e\mu_e\mu_h^2+n_h\mu_h\mu_e^2)B^2}
{(n_h\mu_h+n_e\mu_e)^2 + \mu_h^2\mu_e^2 (n_h-n_e)^2B^2}
\label{eq:rhoxx}
\end{equation}
where $\mu_e$ and $\mu_h$ are the electron and hole mobility respectively, $-e$ is the electron charge.

In the limit $B \rightarrow 0$,
the above equations~\ref{eq:rhoxy}-\ref{eq:rhoxx}
give $\rho_{xx}$ 
and $K_H= \lim_{B \rightarrow 0} \rho_{xy}/B$ as functions of
the parameters $\muchem$, $\mu_e$, $\mu_h$ and $T$.

To explore the role of disorder, 
we introduce the probability $P(V)dV$ of finding the local electronic potential within a range $dV$ about $V$. The disorder is assumed to have a Gaussian form: $P(V) = 1/\sqrt{2\pi s^2} \exp (-V^2 /2s^2)$. The disordered electronic DOS is then given by $D_e(E)= \int_{-\infty}^{E} D_1(E-V) P(V) dV$. A similar formula holds for $D_h$.
%


\end{document}